\documentclass[superscriptaddress,twocolumn,amsmath,amssymb,aps,pra]{revtex4}

\usepackage{amsfonts}
\usepackage{amssymb}
\usepackage{color}
\usepackage{amsmath}
\usepackage{graphicx}
\usepackage{subfigure}
\usepackage{txfonts}
\usepackage{bm}
\usepackage{ulem}

\begin{document}

\title{Determination of the critical exponents in dissipative phase transitions: Coherent anomaly approach }
\author{Jiasen Jin}
\thanks{These two authors contributed equally to the work}
\affiliation{School of Physics, Dalian University of Technology, Dalian 116024, China}

\author{Wen-Bin He}
\thanks{These two authors contributed equally to the work}
\affiliation{Beijing Computational Science Research Center, Beijing 100193, China}
\affiliation{The Abdus Salam International Center for Theoretical Physics, Strada Costiera 11, 34151 Trieste, Italy.}

\author{Fernando Iemini}
\affiliation{Instituto de F{\' i}sica, Universidade Federal Fluminense, 24210-346 Niter{\' o}i, Brazil}
\affiliation{The Abdus Salam International Center for Theoretical Physics, Strada Costiera 11, 34151 Trieste, Italy.}

\author{Diego Ferreira}
\affiliation{Departamento de F{\' i}sica  - Universidade Federal de Minas Gerais, 31270-901, Belo Horizonte - MG - Brazil}

\author{Ying-Dan Wang}
\affiliation{CAS Key Laboratory of Theoretical Physics, Institute of Theoretical Physics,
		Chinese Academy of Sciences, P.O. Box 2735, Beijing 100190, China}
\affiliation{School of Physical Sciences, University of Chinese Academy of Sciences, No.19A Yuquan Road, Beijing 100049, China}
\affiliation{Synergetic Innovation Center for Quantum Effects and Applications, Hunan Normal University, Changsha 410081, China}

\author{Stefano Chesi}
\email{stefano.chesi@csrc.ac.cn}
\affiliation{Beijing Computational Science Research Center, Beijing 100193, China}
\affiliation{Department of Physics, Beijing Normal University, Beijing 100875, China}
\affiliation{The Abdus Salam International Center for Theoretical Physics, Strada Costiera 11, 34151 Trieste, Italy.}

\author{Rosario Fazio}
\email{fazio@ictp.it}
\affiliation{The Abdus Salam International Center for Theoretical Physics, Strada Costiera 11, 34151 Trieste, Italy.}
\affiliation{Dipartimento di Fisica, Universit{\`a} di Napoli "Federico II", Monte S. Angelo, I-80126 Napoli, Italy.}
\affiliation{Beijing Computational Science Research Center, Beijing 100193, China}


\begin{abstract}
We propose a generalization of the coherent anomaly method to extract the critical exponents of a  phase transition occurring in the steady-state of an open quantum
many-body system. The method, originally developed by Suzuki [J. Phys. Soc. Jpn. {\bf 55}, 4205 (1986)] for equilibrium systems, is based on the scaling properties of  the singularity in the
response functions determined through cluster mean-field calculations. We apply this method to the dissipative transverse-field Ising model and the dissipative XYZ model in two dimensions obtaining convergent results already with small clusters.
\end{abstract}

\date{\today}

\maketitle

\section{Introduction}
Phase transitions associated with spontaneous symmetry breaking is a central topic in modern science, appearing in the most diverse situations, both in and out-of equilibrium~\cite{Sachdev,Goldenfeld}. One of the major challenges in describing the critical behaviour of a system on the verge of a transition has  always been the determination of the critical exponents. Several different powerful analytical and numerical methods were elaborated for this purpose~\cite{Domb} which, however, are mostly concerned with equilibrium conditions.

Aim of this paper is to discuss a method that seems particularly suited to determine critical exponents associated to non-equilibrium phase transitions in open quantum many-body system. This question has attracted increasing attention thanks to remarkable breakthroughs in manipulating many-body systems
coupled to an external environment. Steady-state phase transitions have been observed in various experimental platforms as, for example, circuit
QED arrays \cite{fitzpatrick2017,collodo2019}, cold atomic systems in cavities~\cite{ritsch2013} or subject to losses~\cite{tomita2017}, and Rydberg
atom ensembles~\cite{ding2020}. The level of control in these different platforms has reached a stage such that they can be considered in all respects open-system quantum simulators.
An overview of the experimental and theoretical activities can be found in the recent reviews~\cite{houck2012,Sieberer,hartmann2016,noh2017,carusotto2020}.

In all the situations mentioned above, the dynamics is well described by a Lindblad master equation~\cite{breuer} for the reduced density matrix $\rho (t)$ of the system
($\hbar=1$ hereinafter)
\begin{equation}
\dot{\rho}(t)=-i[\hat{H},\rho(t)]+\sum_{\alpha}{{\cal D}_{\alpha}[\rho(t)]},
\label{ME}
\end{equation}
where $\hat{H}$ is the Hamiltonian of the systems under consideration (cavity array, optical lattice, $\dots$), and the superoperator ${\cal D}_{\alpha}[\rho(t)]=
\hat{L}_{\alpha}\rho(t)\hat{L}_{\alpha}^\dagger-\frac{1}{2}\{\hat{L}_{\alpha}^\dagger \hat{L}_{\alpha},\rho(t)\}$, with $\{\cdot,\cdot\}$ being the
anti-commutator,  describes the incoherent dissipation processes induced by the coupling to a memory-less environment. The form of the Lindblad operators
$\hat{L}_{\alpha}$ depends on the process involved in the loss of coherence. The non-equilibrium phase transitions we are referring to take place in the steady state ($t \to \infty$), when the system may undergo spontaneous symmetry breaking. The steady-state phase diagram in many-body open systems governed by Eq.~(\ref{ME}) was analyzed extensively in the recent literature, showing a complexity and
a variety of phases that do not have a counterpart in equilibrium (see~\cite{diehl2010,lee2011,poletti2013,sieberer2013,lee2013,jin2013,marcuzzi2014,
finazzi2015,weimer2015,Marino2016,jin2016,schiro2016,maghrebi2016,wilson2016,zamora2017,kshetrimayum2017,biella2017,rota2019,carollo2019,young2020,
scarlatella2020} and references therein).

In driven-dissipative systems, besides the phase diagram, also the universality class of the transition may change. This question has been studied in several different
models, see e.g.~\cite{marcuzzi2014,maghrebi2016,sieberer2013,carollo2019,young2020,tauber2014,Marino2016}. Therefore it is of great significance to
investigate the exact critical exponents associated with a steady-state phase transition. In particular, it would be very important to have reliable numerical methods to
determine them.  Various powerful methods have been developed to simulate quantum many-body systems~\cite{finazzi2015,kshetrimayum2017,zwolak2004,
daley2014,werner2016,vicentini2019,Nagy2019,Yoshioka2019,hartmann2019,kilda2020,keever2020,weimer2019,Paeckel2019}. It is however important to stress that, from the numerics perspective, directly
accessing the critical properties of the steady-state may be rather demanding. Furthermore, except very few exceptions~\cite{carollo2019}, these phenomena
occur in two (or higher) dimensions, where the numerics becomes considerably more complex. It is therefore desirable to have a method to extract critical exponents that may require moderate resources to reach good accuracy.

With this goal in mind, in this work we apply the so-called Coherent Anomaly Method~\cite{suzuki1986,suzuki1987,katori1987} (CAM) to the steady-state phase
transitions of open quantum many-body systems. The method is based on a combination of Cluster-Mean Field (CMF) approximation with finite-size scaling
of its outcomes. By analyzing the singularities of the CMF response functions, we are able to extract the critical points
and exponents of continuous steady-state phase transitions. The CAM has already been successfully applied in a diversity of equilibrium systems including the Ising model \cite{suzuki1986,suzuki1987,katori1987}, transverse-field Ising model \cite{Nonomura1992} and antiferromagnetic $XXZ$-chain \cite{Nonomura1993}. Moreover, the application of CAM in classical non-equilibrium model has also been reported \cite{dickman2002,park2005}. In particular, for the one-dimensional driven pair contact process with diffusion, the estimates of critical exponents through the CAM analysis are in excellent accord with simulation results \cite{park2005}. Here we explore the power of CAM in accessing the critical exponent in dissipative phase transitions of quantum systems. We tackle this problem by testing the method in two paradigmatic models. In particular, in one case there are solid theoretical predictions that guide our analysis. Additional studies, for other cases, are needed in order to test how powerful is CAM for many-body open quantum systems.

The paper is organized as follows. In the next section, we first briefly review the CMF approach to the dissipative system. We then illustrate the idea of the CAM by considering the magnetic susceptibility and show how to access the critical exponent of dissipative phase transition through a series of CMF results. In Sec. \ref{sec_models}, we introduce two models, namely the dissipative transverse-field Ising model and the dissipative XYZ model on two-dimensional square lattice, as examples to investigate the performance of CAM. The details of numerical simulation are explained as well (Sec. \ref{sec_numerical}). In Sec. \ref{sec_results}, the critical exponent and critical point that extracted from CAM for both models are presented. We compare the CAM results with the classical critical exponent and discuss the stability of the CAM results through different choices of clusters. Finally, we summarize in Sec. \ref{sec_summary}.

\begin{figure*}
	\centering
	\includegraphics[width=0.9\linewidth]{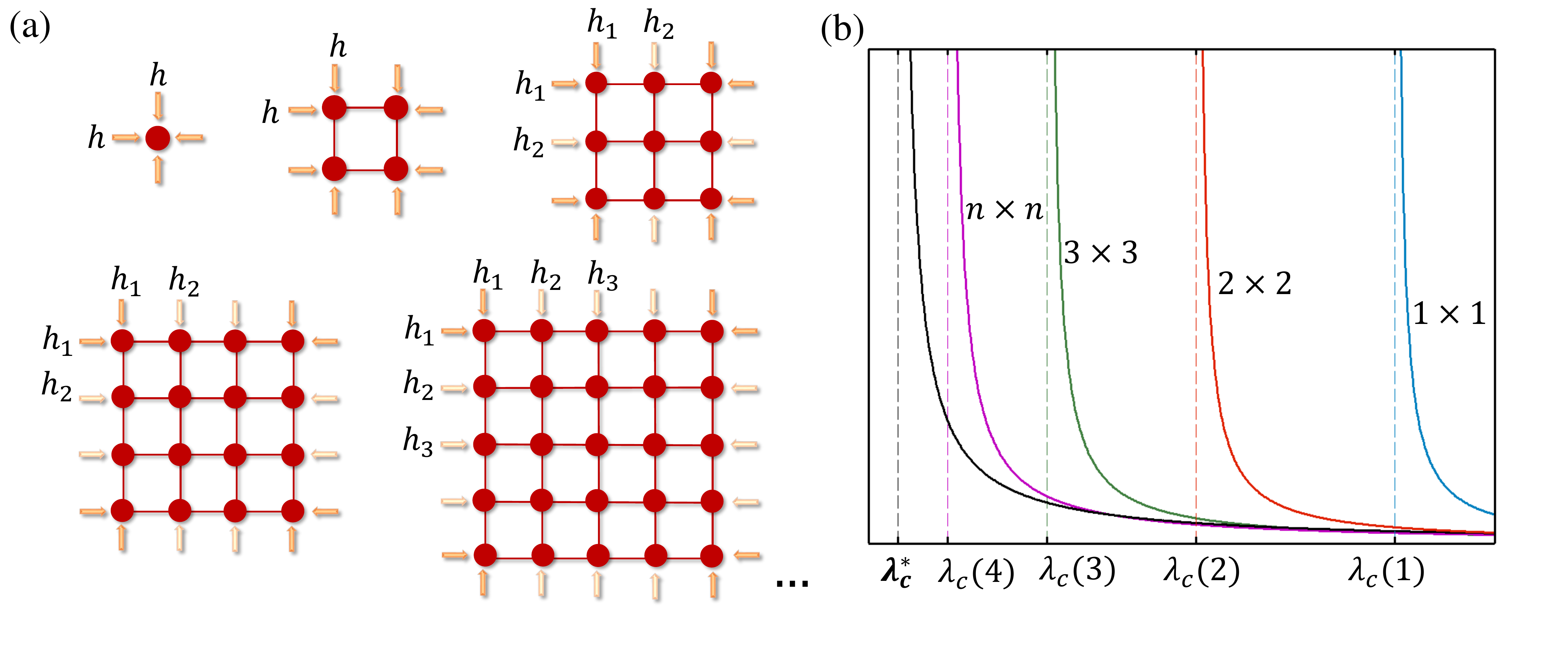}
	\caption{(Color online) (a) Series of isotropic clusters for a systematic CMF approximations. In the CMF simulation, the global density matrix of the whole
	lattice  is  factorized as the tensor product of density matrices of identical finite-size clusters.  The interactions inside the cluster (solid lines) are treated
	exactly while each boundary site is influenced by effective fields $h_j$ (arrows) from the adjacent clusters, which are determined by a  self-consistent condition.
	(b) Main idea of the coherent anomaly method. The susceptibilities for a set of CMF approximations diverge at the critical points $\lambda_c(n)$ ($n=1,2,...$).  While all such divergences are governed by the
 classical exponent $\gamma^{\text{mf}}$, the amplitudes $\chi_0(n)$ [defined in Eq.~(\ref{chi_mf})] become anomalously large as $\lambda_c(n)$ approaches the true critical point $\lambda_c^*$. The divergence of $\chi_0(n)$ gives the true critical exponent $\gamma^*$, according to Eq. (\ref{fL}).}
\label{Fig1}
\end{figure*}

\section{Coherent Anomaly Method}
\label{sec_cam}
The CAM method is based on a scaling analysis of mean-field results. To set the stage, we start with a
brief review of the CMF approximation applied to driven-dissipative systems. The CMF approach is based on the factorization of the global density matrix
$\rho(t)=\bigotimes_C{\rho_{C}(t)}$. The clusters in the factorization are assumed to be identical. As a consequence, Eq. (\ref{ME}) is decoupled into the following master
equation with respect to the cluster $C$,
\begin{equation}
\dot{\rho}_{C}(t)=-i[\hat{H}_{\text{CMF}}(\lambda,t),\rho_{C}(t)]+\sum_{j\in C}{{\cal D}_j[\rho_{C}(t)]},
\label{ME_CMF}
\end{equation}
where $\lambda$ is a driving parameter and we supposed that dissipation acts locally on the sites $j$ of the lattice. Considering $\hat H = \sum_{\alpha} \sum_{\langle j,k \rangle} \hat{H}_{jk}^{(\alpha)}$ (which includes various nearest-neighbor interactions of type $\alpha$, with factorized form $\hat{H}^{(\alpha)}_{jk}=\hat{H}^{(\alpha)}_{j}\otimes \hat{H}^{(\alpha)}_{k}$), the CMF Hamiltonian is given by $\hat{H}_{\text{CMF}}(\lambda,t)=\hat{H}_{C}(\lambda)+\hat{H}_{\partial C}(\lambda,t)$,
where $\hat{H}_C(\lambda)=\sum_\alpha\sum_{\langle j,k\rangle\in C}{\hat{H}^{(\alpha)}_{jk}}$ describes the interactions inside the cluster and $\hat{H}_{\partial C}(\lambda,t)=
\sum_{\alpha}\sum_{j\in\partial C}{h^{(\alpha)}_j(t)\hat{H}^{(\alpha)}_j(\lambda)}$ describes the interaction of sites on the boundary with the adjacent clusters. The time-dependent
effective fields are given by $h^{(\alpha)}_j(t) = \sum^\prime_{k} \text{Tr}[\hat{H}^{(\alpha)}_{k} \rho_C(t)]$, where the prime indicates sites $k\in C$ corresponding to neighbors of $j$. Due to translational invariance, the long-time limit asymptotic solution of Eq. (\ref{ME_CMF}) is determined by the following self-consistent condition
\begin{equation}
h^{(\alpha)}_j(t\rightarrow\infty)=\text{Tr}[\hat{h}^{(\alpha)}_j \rho_C(t\rightarrow\infty)],
\label{scc}
\end{equation}
where $\hat{h}^{(\alpha)}_j=  \sum^\prime_{k} \hat{H}^{(\alpha)}_{k} $ are the local effective field operators. \color{black}
A sketch of the cluster mean-field approach is illustrated in Fig.~\ref{Fig1}.
From Fig.~\ref{Fig1}(a), one can see that for the standard (single-site) mean-field approximation the effective field governs the coherent evolution, while $\hat{H}_C(\lambda)$
is absent; as the size of cluster becomes larger, the effective fields acting on the boundary sites play a less dominant role in $\hat{H}_{\text{CMF}}$, and the quantum correlations
embedded in $\hat{H}_C(\lambda)$ become more significant. Through systematically enlarging the cluster size, the gradual inclusion of correlations and the self-consistent
conditions results in stronger singular behaviors of the order parameter and response function. Such coherently enhanced singularities are intimately related to the intrinsic
fluctuations embedded in the system. Specifically, the increased fluctuations in the cluster leads to the divergence of the amplitude of the response function as the size of cluster
goes to infinity. The main idea behind the CAM~\cite{suzuki1986,suzuki1987,katori1987} is to analyze how the mean-field results change with the size of the cluster, which will
provide information of the (non mean-field) exponents as well as the exact boundaries.
The choice of the series of clusters in the CAM should guarantee that, as the cluster size $L$ is systematically enlarged, the set of critical points $\lambda_c(L)$ obtained from successive
CMF calculations should asymptotically approach the true critical point $\lambda^*$, i.e.,
 \begin{equation}
\lim_{L\rightarrow\infty}{\lambda_c(L)}=\lambda^*.
\label{systematic}
\end{equation}
This is a key point of CAM because some choices of
the cluster series may not satisfy Eq.~(\ref{systematic}). For example, in a two-dimensional square lattice, the CMF analysis using a series of plaquette clusters of size
$L = l_x\times l_y$ may not approach the true critical point if $l_x\gg l_y$, because of the large anisotropy; thus we will choose the square plaquette of size $L$ with
$l_x=l_y=l$ to implement the CAM.

In order to illustrate the idea of CAM~\cite{suzuki1986,suzuki1987,katori1987} we consider, for concreteness,  the magnetic susceptibility, also in the light of the models we are going to
study. In the presence of a small external magnetic field $\delta$
 and taking the self-consistency condition
into account, the magnetization in the $x$ direction on the corner-site of a
cluster may be written as
\begin{equation}
\langle \hat{\sigma}^x(\lambda)\rangle_{\text{ss}} = \chi_L^{\text{cl}}(\lambda)\delta + F_L(\lambda)\langle\hat{\sigma}^x(\lambda)\rangle_{\text{ss}},
\label{mag_cmf}
\end{equation}
where the two terms on right-hand side denote the magnetic responses to the external and self-consistent fields, respectively. The subscript `ss' denotes the steady-state. $\chi_L^{\text{cl}}(\lambda)$ is the magnetic susceptibility of the cluster
without the self-consistent field and $F_L(\lambda)$ is Kubo's canonical correlation. Consequently, the magnetic susceptibility as a function of $\lambda$ is given by
$\chi(\lambda)=\chi^{\text{cl}}_{L}(\lambda)/\left[ 1-F_L(\lambda)\right]$. The critical point $\lambda_c(L)$ is determined by $1-F_L(\lambda)=0$.
In the CMF approximation, the behavior of the response function near the transition point is characterized by the classical critical exponent according to the Ginzburg-Landau
theory. Thus, in the vicinity of $\lambda_c(L)$ the magnetic susceptibility can be approximated by
\begin{equation}
\chi(\lambda)\approx \chi_0(L)\left[\frac{\lambda-\lambda_c(L)}{\lambda_c(L)}\right]^{-\gamma^{\text{mf}}},
\label{chi_mf}
\end{equation}
where $\gamma^{\text{mf}}=1$ is the classical mean-field critical exponent of the susceptibility.
The amplitude in Eq. (\ref{chi_mf}) takes the form  $\chi_0(L)\approx\chi_L^{\text{cl}} (\lambda_c(L))/[-\lambda_c(L)F_L'(\lambda_c(L))]$. It reflects the fluctuations stemming
from the inclusion of short-range correlations in the relevant cluster and should be anomalously large as $\lambda_c(L)$ goes to $\lambda^*$ (or as $L\rightarrow\infty$).
The dependence of $\chi_0(L)$ on $\lambda_c(L)$ is called {\it coherent anomaly}, and carries the intrinsic information of the true critical behavior through its large-$L$ asymptotic dependence~\cite{suzuki1986},
\begin{equation}
\chi_0(L) \approx c_0\left[\lambda_c(L)-\lambda^*\right]^{-(\gamma^*-\gamma^{\text{mf}})},
\label{fL}
\end{equation}
where $\gamma^*$ is the true critical exponent  and $c_0$ a prefactor. In Eq.~(\ref{fL}), there are three unknown quantities, $c_0,\lambda^*,\gamma^*$, which demands at
least the results of three different cluster sizes. As a consequence, the true critical exponent can be estimated as
\begin{equation}
\gamma^*= \gamma^{\text{mf}} + \frac{\log{\left[\chi_0(j)/\chi_0(k)\right]}}{\log{\left\{\left[\lambda_c(k)-\lambda^*\right]/\left[\lambda_c(j)-\lambda^*\right]\right\}}},
\label{gammastar}
\end{equation}
with $\{\chi_0(j),\lambda_c(j)\}$ being a set of amplitudes and critical points obtained by CMF approximations, with at least three successive systematically enlarging clusters.

The CAM procedure, in practice, goes along the following steps: (i) compute the response function $\chi(\lambda)$ at various levels of CMF approximation with systematically
enlarging clusters; (ii) extract the coefficients and critical points $\{\chi_0(L),\lambda_c(L)\}$ for the various CMF approximation with Eq. (\ref{chi_mf}); (iii) fit the dependence of $\chi_0(L)$ on
$\lambda_c(L)$ with Eq.~(\ref{fL}), to access the true critical point and exponent.

We finish this section by emphasizing that the coherent anomaly relation in Eq. (\ref{fL}) is derived in a rather general way. It doesn't make any reference to the underlying nature of the system or whether the system is equilibrium/non-equilibrium or classical/quantum. The validity of the application of CAM depends only on the convergence of the properties of scaling functions. Namely, the CAM can be applied to the system that the true critical point can be reached when the cluster of system asymptotically tends to be infinite. As shown in Refs. \cite{suzuki1986,park2005}, CAM obtained good results of the phase transitions in equilibrium and non-equilibrium systems. Based on above arguments, we apply CAM to study the critical exponents in dissipative phase transitions of quantum systems.

\section{The Models}
\label{sec_models}
To investigate explicitly the performance of CAM, we will consider two models of dissipative quantum many-body systems. The first one is the
spin-1/2 dissipative TFI model on the square lattice. The Hamiltonian governing the coherent evolution is the following,
\begin{eqnarray}
\hat{H}^{\text{TFI}} &=&-\frac{V}{4}\sum_{\left\langle j,k \right\rangle}{\hat{\sigma}^x_j\hat{\sigma}^x_k} +\frac{g}{2} \sum_k{\hat{\sigma}^z_k},
\end{eqnarray}
where $\hat{\sigma}_j^\alpha$ ($\alpha=x,y,z$) denote the spin-1/2 Pauli matrices on the $j$-th site and $\langle j,k\rangle$ denotes the nearest-neighbor spins. The first term in $\hat{H}^{\text{TFI}}$ represents Ising interactions of strength $V$ between nearest-neighbor sites, while the second term accounts for the local transverse magnetic field along the $z$ direction,  with amplitude $g$.

As the second example, we will investigate the two-dimensional dissipative spin-1/2 XYZ model on the square lattice. The Hamiltonian in Eq. (1) for this model is given by
\begin{equation}
\hat{H}^{\text{XYZ}}=\sum_{\langle j,k\rangle}{\left(J_x\hat{\sigma}^x_j\hat{\sigma}^x_k+J_y\hat{\sigma}^y_j\hat{\sigma}^y_k+J_z\hat{\sigma}^z_j\hat{\sigma}^z_k\right)},
\label{Hxyz}
\end{equation}
where $J_\alpha$ ($\alpha=x,y,z$) is the coupling strength.

For both models, we consider an incoherent dissipative process acting locally on each site which tends to flip the spin down along the $z$-direction, namely
$L_j=\sqrt{\Gamma}(\hat{\sigma}_j^x-i\hat{\sigma}^y_j)/2 =\sqrt{\Gamma} \hat{\sigma}_{j}^{-}$ in Eq. (\ref{ME}), where $\Gamma$ is the decay rate.
The master equation of both models presents a $\mathbb{Z}_2$ symmetry, namely the master equation is invariant under a $\pi$-rotation about the $z$-direction, $(\hat{\sigma}_{j}^x,\hat{\sigma}_{j}^y)\rightarrow(-\hat{\sigma}_{j}^x,-\hat{\sigma}_{j}^y)$. We define the order parameter as the averaged steady-state magnetization per site
${\cal O}=\langle\hat{\sigma}^{x,y}\rangle_{\text{ss}}=\frac{1}{L}\sum_{j=1}^{L}{\text{tr}\left(\rho_{\text{ss}}\hat{\sigma}^{x,y}_j\right)}$, wher $j$ labels the site in the lattice of size $L$ and $\rho_{\text{ss}}$ is the steady-state density matrix. The nonzero order parameter indicates appearance of the ordered phase.

\begin{figure}[htp]
  \includegraphics[width=1\linewidth]{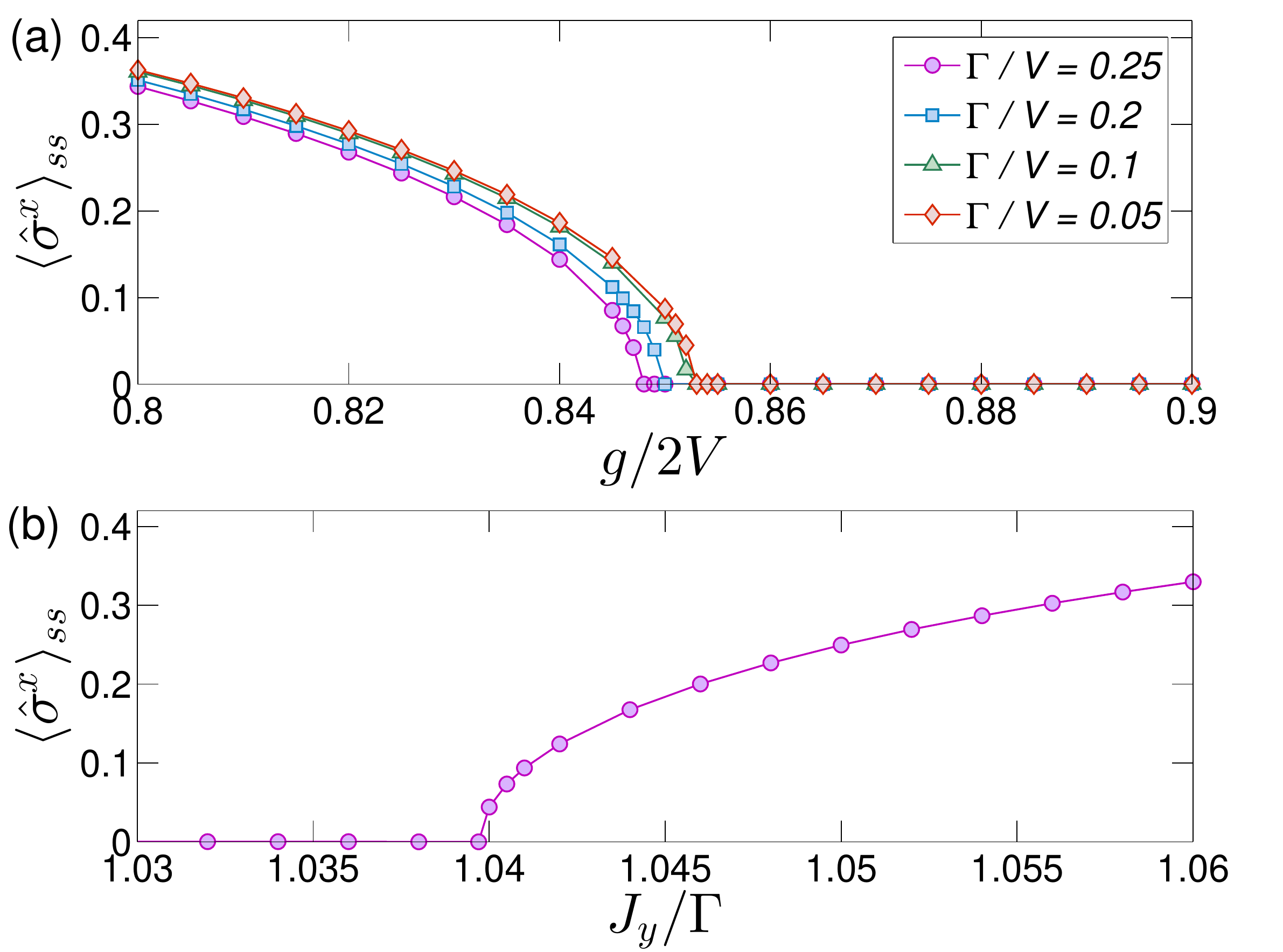}
  \caption{The cluster mean-field order parameter $\langle\hat{\sigma}^{x}\rangle_{\text{ss}}$ for (a) the dissipative TFI model as a function of $g/2V$ for the cases of $\Gamma/V=0.05$, $0.1$, $0.2$ and $0.25$, (b) XYZ model, whose Hamiltonian is given by Eq. (\ref{Hxyz}), as function of $J_y/\Gamma$ for $J_x / \Gamma= 0.9$ and $J_z / \Gamma = 1$. }
  \label{magn}
\end{figure}

By means of Keldysh formalism \cite{overbeck2017} and cluster mean-field
approximation \cite{jin2018}, a continuous phase transition from the disordered paramagnetic (PM) phase ($\langle\hat{\sigma}^{x}\rangle,\langle\hat{\sigma}^{y}\rangle = 0$),
to the ferromagnetic (FM) phase ($\langle\hat{\sigma}^{x}\rangle,\langle\hat{\sigma}^{y}\rangle\neq0$) has been predicted for the dissipative TFI model. In Fig. \ref{magn}(a), we show the order parameter as a function of the strength of the transverse field by means of $3\times3$ CMF simulation.

Meanwhile, in the dissipative XYZ model, a steady-state phase transition breaking the $\mathbb{Z}_2$ symmetry has been predicted \cite{finazzi2015,biella2018,kshetrimayum2017,huybrechts2019,rota2017}. For $J_x=J_y$, the magnetization along $z$-direction is a conserved quantity and there is nothing counteracts to the dissipation, thus leads all the spins down to the $z$-direction. The steady state of the system is in the disordered PM phase. However, the anisotropic coupling $J_x\ne J_y$ in the $x$-$y$ plane of the Hamiltonian may induce an effective field to each spin. The competition between the procession of the spin around the effective field and the decay leads to a phase transition from the disordered PM to ordered FM phase. Again, the steady-state phase transition can be revealed by the order parameter as shown in Fig. \ref{magn}(b). Moreover, the steady-state phase diagram has been predicted by CMF approaches \cite{jin2016}.

The steady-state phase transition in both two dissipative models can also be captured by the singular behavior of the magnetic susceptibility. The susceptibility is defined as the linear response of the magnetization to the small probing field. The probing field modifies the CMF Hamiltonian according to
\begin{equation}
\hat{H}_{\text{CMF}}\rightarrow\hat{H}_{\text{CMF}}+\delta_\beta\sum_k{\hat{\sigma}^\beta_k},{\forall k}
\label{Hcmf_ext}
\end{equation}
where $\delta_\beta$ is the strength of the external driving field along $\beta$-direction ($\beta = x, y$). The magnetic susceptibility is a four-component tensor. Each component characterizes the response magnetization along $\alpha$-direction to the probing field along $\beta$-direction and is given by
\begin{equation}
\chi_{\alpha\beta} =\lim_{\delta_{\beta}\rightarrow 0}{ \frac{\partial \langle\sigma^\alpha \rangle_{\text{ss}}  }{\partial \delta_{\beta}}},
\label{chiab}
\end{equation}
where $\alpha,\beta=\{x,y\}$ denote the directions of external field and response magnetization, respectively.
In the following, we will concentrate on the modulus of the $\chi_{xx}$ component of dissipative TFI model while $\chi_{yx}$ for disspative XYZ model.

\begin{table*}
\centering
\caption{The coefficients, critical points and critical exponents of steady-state phase transition in the dissipative TFI model extracted from the CAM with different groups of clusters. The errors indicate $95\%$ confidence intervals, obtained from the estimate by least-squares fitting.}
\begin{tabular}{p{1.2cm}|p{1.8cm}p{1.8cm}p{1.8cm}|p{1.8cm}p{1.4cm}p{1.8cm}|p{1.8cm}p{1.4cm}p{1.8cm}cccccccccc}
\hline
\hline
&\multicolumn{3}{c|}{($1\times1$, $2\times2$, $3\times3$, $4\times4$)} & \multicolumn{3}{c|}{($1\times1$, $2\times2$, $3\times3$)} &\multicolumn{3}{c}{($2\times2$, $3\times3$, $4\times4$)} \\
\hline
$\Gamma/V$&$c_0$&$g^*/2V$&$\gamma^*$&$c_0$&$g^*/2V$&$\gamma^*$&$c_0$&$g^*/2V$&$\gamma^*$ \\
\hline
0.1& 0.059(18)& 0.788(43)& 1.33(20)& 0.058& 0.776&   1.37& 0.063& 0.794& 1.30\\
0.2& 0.119(15)& 0.787(16)& 1.335(81)& 0.117& 0.783& 1.35& 0.122& 0.790& 1.32\\
0.05& - & - & - &   0.028&0.768& 1.34& - & - & - \\
0.25&  - & - & - & 0.146&0.775& 1.37 & - & - & - \\
\hline
\hline
\end{tabular}
\label{cam_stra}
\end{table*}

\subsection{Numerical simulations}
\label{sec_numerical}
In this subsection, we will show the way for accessing the steady-state magnetization and susceptibility in our specific models. In the CMF approximations, one need to solve the CMF master equation Eq. (2). For large clusters, this is demanding since the dimension of Hilbert space increases exponentially with the size increasing. In order to avoid solving the full master equation regarding the density matrix of size $2^L\times 2^L$, one can combine the idea of CMF and quantum trajectory (QT) simulation.  In this paper, for $l\le3$, we evaluate the master equation~(\ref{ME_CMF}) to long times through direct Runge-Kutta integration. Instead, for $l=4$ we rely on QT simulations (with the number of trajectories being 500). The details of CMF QT simulation are discussed in Refs. \cite{jin2016,daley2014}.

The steps of accessing the steady-state susceptibility through QT are the following. We first simulate the time-evolution of the density matrix in the presence of probing field $\delta_x$. Second, we average the temporary magnetization over a time window, in which the system nearly approaches to the steady state, as the steady-state magnetization $\langle\hat{\sigma}^{x,y}\rangle_{\text{ss}}$ . Finally we determine the susceptibility by linear fitting of $\langle\hat{\sigma}^{x,y}\rangle_{\text{ss}}$ to $\delta_x$.

\begin{figure*}[htb]
  \includegraphics[width=0.9\linewidth]{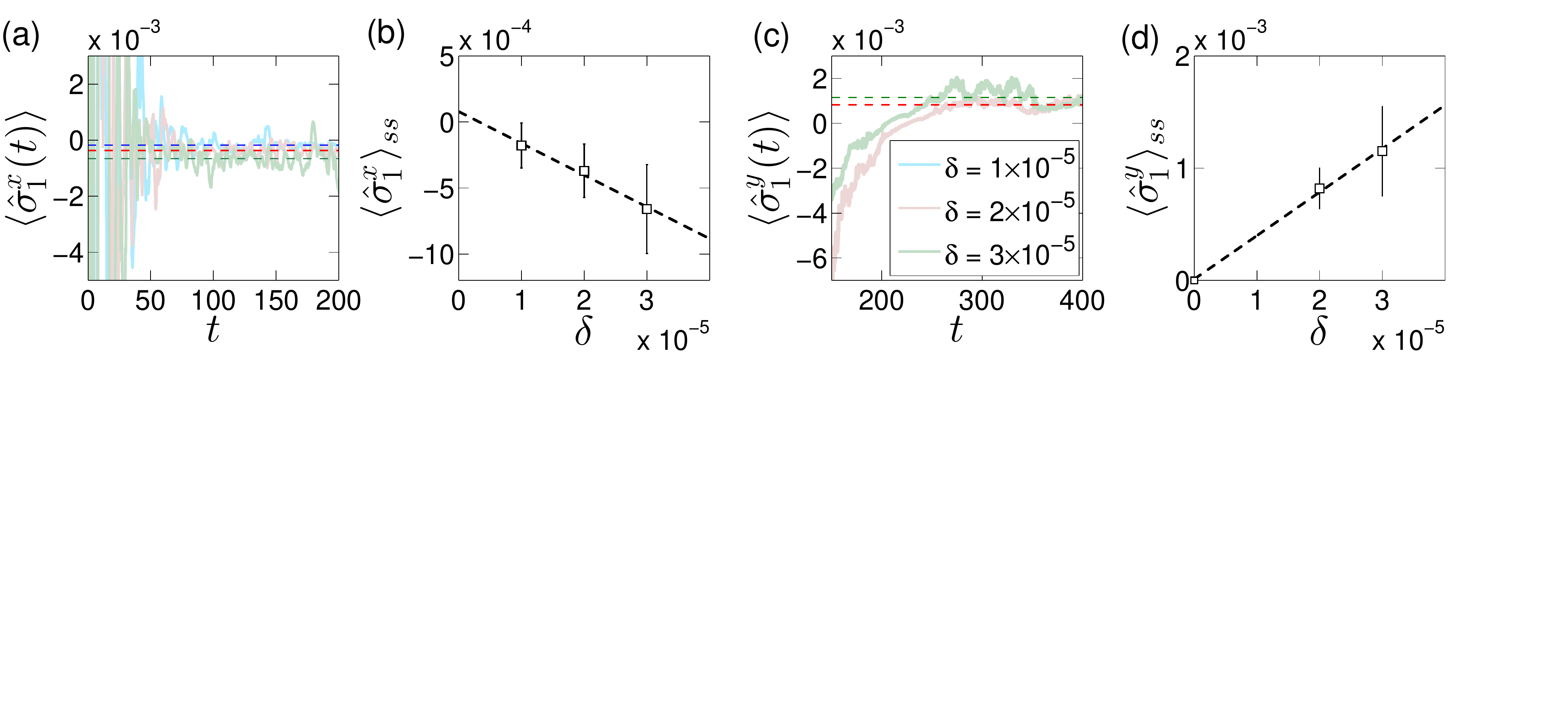}
  \caption{(a) The QT time-evolutions of the magnetization $\langle\sigma_1^x(t)\rangle$, which is the magnetization of the corner site of the $3\times4$ cluster, in the dissipative TFI model for three different probe fields $\delta_x$ (in units of $\Gamma$). The dashed horizontal lines highlight the steady-state values of the magnetization, which is obtained by averaging the temporary magnetization over $t\in{\left[150,200\right]}$. (b) The steady-state magnetization of TFI model $\langle\sigma_1^x\rangle_{\text{ss}}$ for different values of the probing field $\delta_x$. The black dash line is the linear fitting for $\langle\sigma_1^x\rangle_{\text{ss}}$ to $\delta_x$. The errorbars denote the standard deviation in taking the average. The parameters are chosen as $\Gamma/V = 0.2$ and $g/2V = 0.85$. (c) The $4\times4$ quantum trajectory time-evolution of $\langle\sigma^y_1(t)\rangle$ in the presence of probing field along $x$-direction in two-dimensional dissipative XYZ model. The horizontal dashed lines highlight the steady-state magnetizations. (d) The steady-state magnetization of XYZ model $\langle\sigma^y_1\rangle_{\text{ss}}$ versus the probing field $\delta_x$. The squares denote the steady magnetization which is obtained by averaging the temporary magnetization over $t\in[300,400]$, the error bars denote the standard deviation. The black dashed line is the linear fitting. The susceptibility is given by the slope of the straight line.  Other parameters are chosen as  $J_x/\Gamma=0.9$, $J_y/\Gamma = 1.036$ and $J_z/\Gamma = 1$.}
  \label{cam_steadychi}
\end{figure*}

In Fig. \ref{cam_steadychi}(a), we show the QT time-evolution of the magnetization along $x$-direction $\langle\hat{\sigma}_1^x(t)\rangle$ for the corner site in the $3\times4$ cluster of TFI model. The number of trajectories is 500. Although the temporary magnetization fluctuates due to the randomness inherent to the QT method, the system approaches to the steady state at $t>150$. The steady-state magnetization $\langle\hat{\sigma}_1^x\rangle_{\text{ss}}$ is then obtained by averaging over the time interval $t\in\left[150,200\right]$. The linear fits for $\langle\hat{\sigma}_1^x\rangle_{\text{ss}}$ against $\delta$ is shown in Fig. \ref{cam_steadychi}(b). Similarly, in Figs. \ref{cam_steadychi} (c) and (d), the QT time-evolution of the magnetization $\langle\hat{\sigma}_1^y(t)\rangle$ for different $\delta_x$ and the linear fits for the steady-state magnetization in the dissipative XYZ model are shown.

\begin{figure}[h]
  \includegraphics[width=1\linewidth]{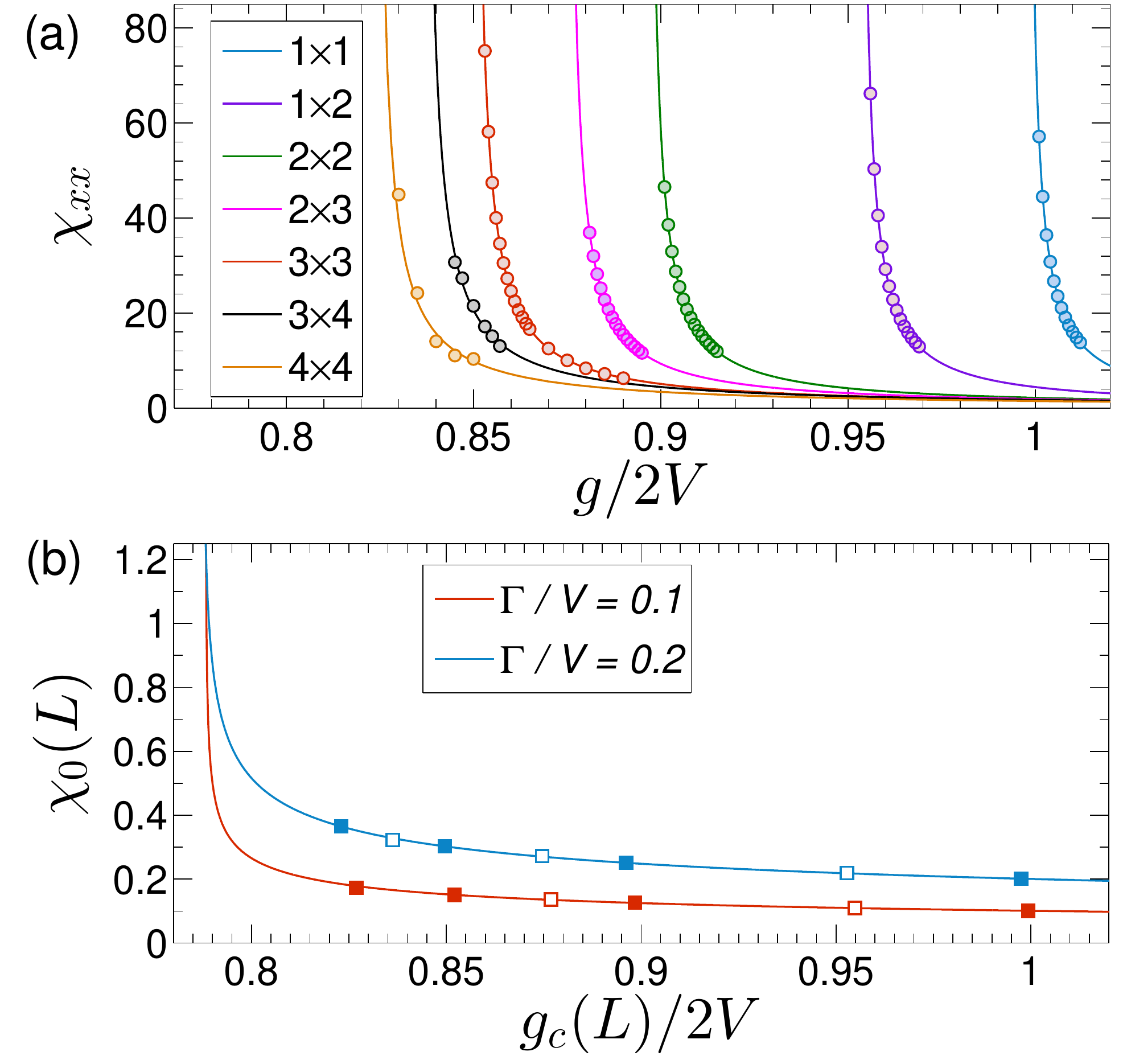}
  \caption{(a) Modulus of $\chi_{xx}$,  computed for various cluster sizes as a function of $g/2V$, with $\Gamma/V=0.2$. (b) Behavior  of $\chi_0(L)$ as a function
  of $g_c(L)/2V$ for $\Gamma/V = 0.1$ (red) and $0.2$ (blue). The solid lines are least-squares fits of Eq.~(\ref{fL}), in which $c_0$, $g^*/2V$ and $\gamma^*$ are determined
  by the $l\times l$ ($l=1,2,3,4$) CMF data (filled squares). The $\chi_0(L)$  and $g_c(L)/2V$ of $1\times2$, $2\times3$ and $3\times4$ CMF data are also presented (empty squares).
  }
  \label{TFI_cam_v1}
\end{figure}

\section{Results}
\label{sec_results}
\subsection{Dissipative TFI model}
\label{sec_results_TFI}
In this subsection, we present the critical exponent and critical point obtained by the CAM for the dissipative TFI model. In Fig.~\ref{TFI_cam_v1}(a) we fix  $\Gamma/V=0.2$ and show the CMF results for $\left|\chi_{xx}\right|$ as function of $g/2V$, with various cluster sizes.
One can see that, for each $L$, the magnetic susceptibility always exhibits a singularity at the critical point $g_c(L)/2V$. Moreover, the critical points shift to the left with increasing cluster
size. The consistent approach of $g_c(L)$ to the true critical point $g^*$ supports choosing $l\times l$ clusters
as a systematic series of approximations. The extracted critical points and corresponding amplitudes $\chi_0(L)$ are shown in Fig.~\ref{TFI_cam_v1}(b).

In order to access the critical point $g^*/2V$ and exponent $\gamma^*$ in the thermodynamic limit, we fit the $\{\chi_0(L),g_c(L)/2V\}$ data of $l\times l$ clusters ($l=1,2,3,4$)
according to Eq.~(\ref{fL}), using the least squares method. The fit is shown by the solid line in Fig.~\ref{TFI_cam_v1}(b). We also find that the CAM analysis is robust against slight anisotropies of the clusters. Although the fit of the coherent anomaly is based on the isotropic clusters (denoted by filled squares), the data of $l\times (l+1)$ clusters (denoted by empty squares) show good agreement.  A similar CAM analysis is implemented for $\Gamma/V = 0.1$. The  set of $\{\chi_0(L),g_c(L)/2V\}$ and the corresponding fitting curve are shown in Fig.~\ref{TFI_cam_v1}(b) as well.

\begin{figure}[h]
  \includegraphics[width=1\linewidth]{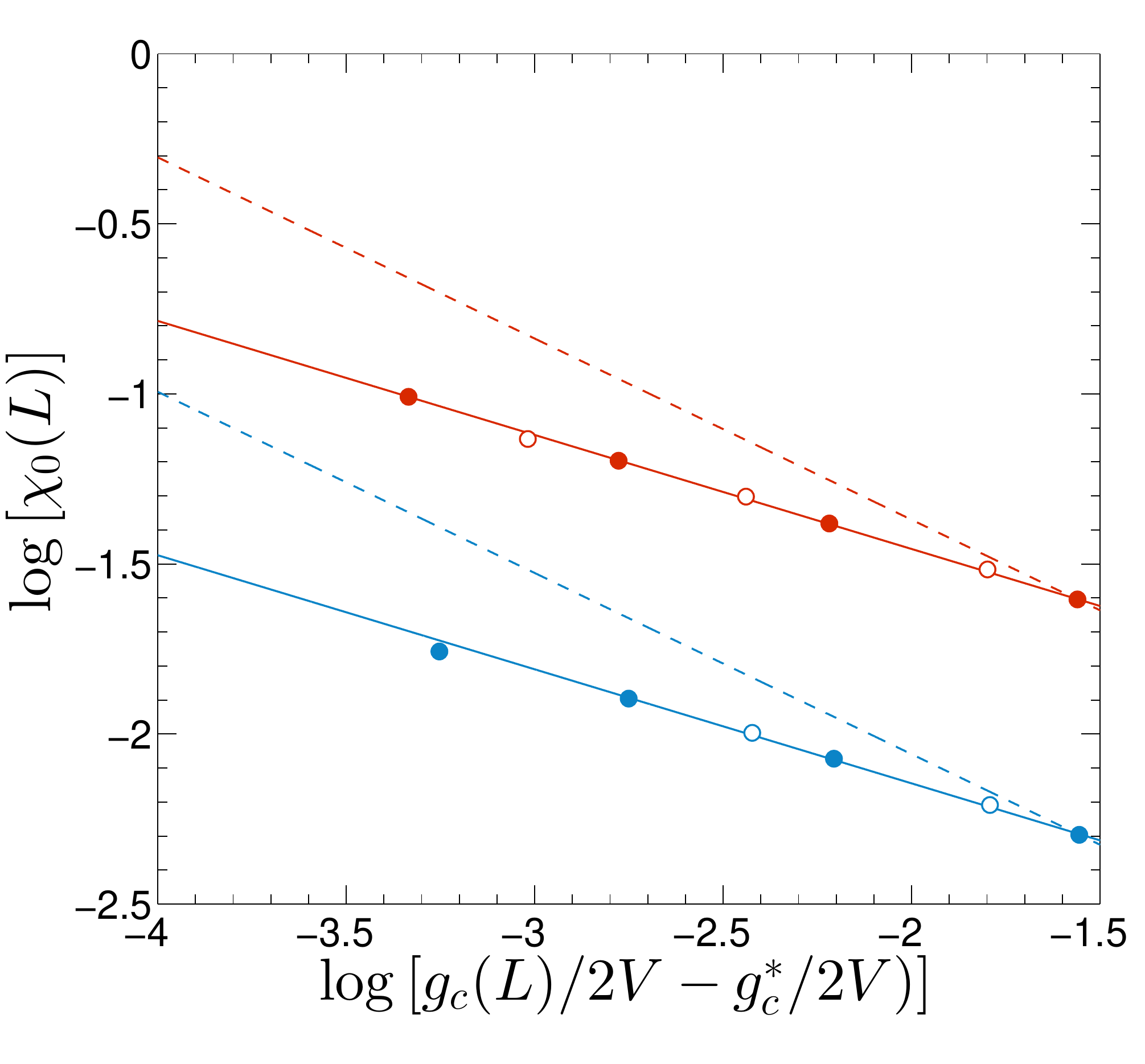}
  \caption{Log-log plot of amplitudes as functions of critical points $g_c(L)$, for $\Gamma/V = 0.1$ (red) and $0.2$ (blue).
  The squares are the amplitudes $\chi_0(L)$. The solid lines use the averaged critical exponents of the dissipative TFI model, $\bar{\gamma}^* = 1.34$. The dashed lines assume the exact critical exponent of the classical 2D Ising model, $\gamma^{\text{Ising}} = 1.75$.
  }
  \label{TFI_loglog}
\end{figure}

For both values of $\Gamma/V$, we obtain a critical exponent $\gamma^*\approx 1.3$. It is interesting to compare this value to Maghrebi {\it et.al.} which, from field-theory arguments, argued that the steady-state phase transition belongs to the universality class of the equilibrium 2D Ising model \cite{maghrebi2016}. Remarkably, our CAM analysis gives a critical exponent significantly smaller than critical exponent of classical 2D Ising model $\gamma^{\text{Ising}}=7/4$. The difference between the two predictions can be appreciated visually in the log-log representation of coherent amplitudes versus critical points, shown in Fig.~\ref{TFI_loglog}. While the numerical data follow an approximate linear relation, the slope is quite distinct from the critical exponent of the classical 2D Ising model (dashed lines).

To test the stability of the CAM analysis, we have adopted two other possible choices of $l\times l$ clusters: $l=1,2,3$ and $l=2,3,4$. In both cases, $c_0,\lambda^*,\gamma^*$ can be extracted by a direct solution of Eq. (\ref{gammastar}). The results are listed in Table~\ref{cam_stra} and indicate a good stability of the CAM results. In particular, the averaged value of $\gamma^*$ over the three choices is $\bar{\gamma}^*\approx 1.34$ for both $\Gamma/V = 0.1$ and $0.2$, with small discrepancies $\Delta\gamma^*=|\gamma^*-\bar{\gamma}^*|/\bar{\gamma}^* \lesssim 3\%$.

On the other hand, the error associated with the limited cluster size, $l\leq 4$,  is difficult to assess. Some insight is provided by the CAM analysis of the classical 2D Ising model. By only considering $l=1,2$ (with the exact critical temperature as additional input), Suzuki obtained  $\gamma^{\text{Ising}}\approx 1.67$ \cite{suzuki1986}. This value is relatively close to the exact result 1.75, suggesting that CAM can yield good estimates of the critical exponents already at small cluster sizes. However, considerable efforts might be necessary to improve the accuracy. Extending the analysis of the classical 2D Ising model to clusters with up to 145 sites ($l \simeq 12$) gives $\gamma^{\text{Ising}}\approx 1.7$ \cite{suzuki1987}.

Additionally, we also extract the critical points and critical exponents for $\Gamma / V = 0.05$ and $0.25$ by CAM involving only the $1\times 1$, $2\times2$, and $3\times 3$ clusters. The results are shown in Tab. \ref{cam_stra}.

\subsection{The dissipative XYZ model}
\label{sec_results_XYZ}
\begin{figure}[htb]
  \includegraphics[width=1\linewidth]{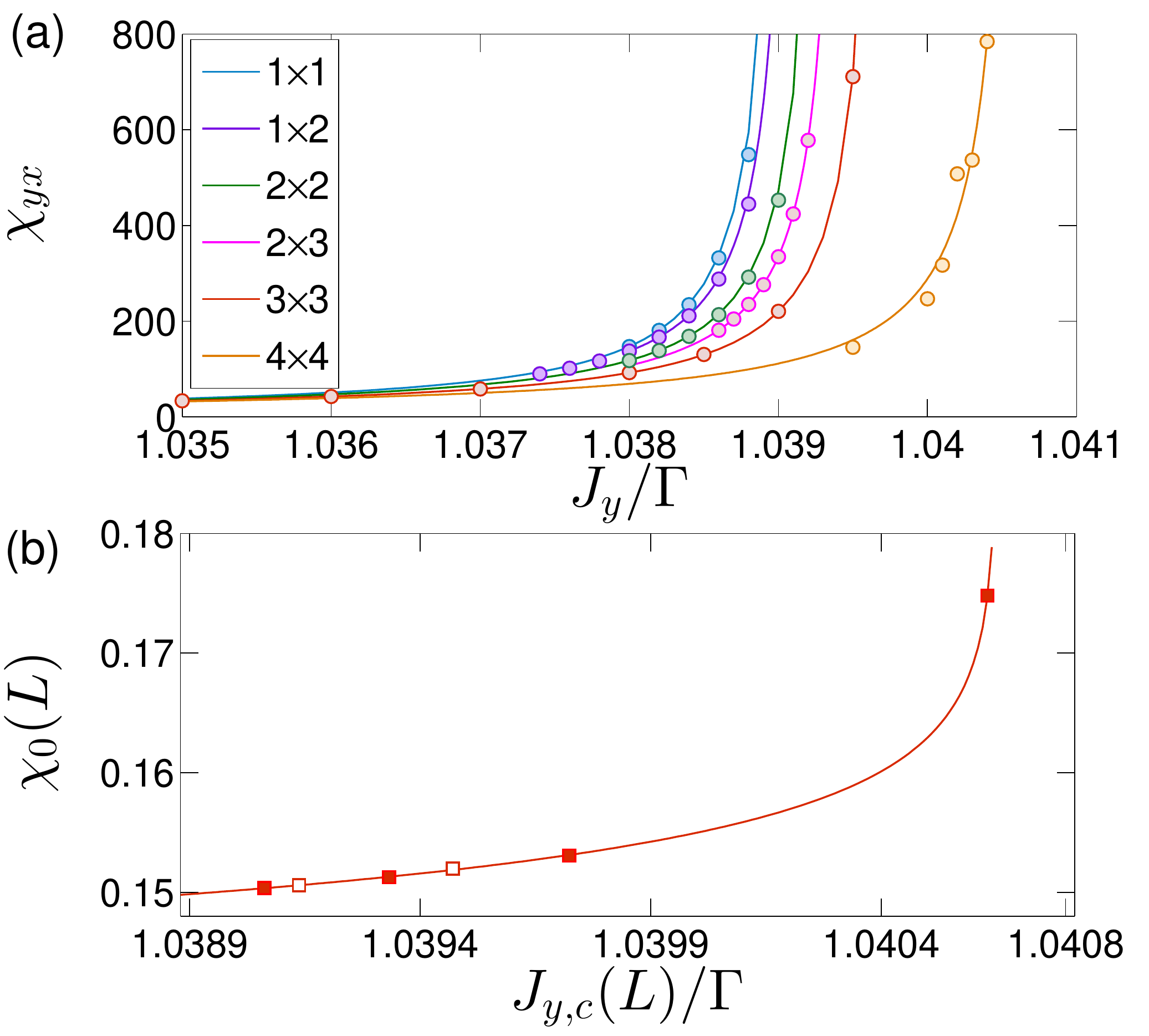}
  \caption{(a) The susceptibility $\chi_{yx}$ as a function of $J_y$ for different clusters. (b)The amplitude $\chi_0(L)$ as a function of $J_{y,c}(L)$. The parameters are chosen as $J_x/\Gamma = 0.9$ and $J_z/\Gamma = 1$. From right to left the squares denote the amplitudes $\chi_0(L)$ for $1\times1$, $1\times2$, $2\times2$, $2\times3$, $3\times3$, and $4\times4$ cluster, respectively.
  }
  \label{XYZ_cam_v1}
\end{figure}

For the dissipative XYZ model, we focus on the critical exponent of the susceptibility component $\chi_{yx}$. In Fig.\ref{XYZ_cam_v1} (a), we show the CMF results of the steady-state susceptibility with different clusters for $J_x/\Gamma = 0.9$. The amplitude $\chi_0(L)$ and the critical point $J_{y,c}(L)/\Gamma$ for various clusters are extracted according to Eq.(\ref{chi_mf}) and shown in Fig.\ref{XYZ_cam_v1}(b).
The CAM is then implemented up on the three choices of clusters. The extracted coefficient, critical exponent and critical point are listed in Tab. \ref{cam_XYZ1}. The critical point $J_{y}^*/\Gamma$ obtained from the CAM analysis is consistent with the previous results through other approaches \cite{jin2016}. The CAM analysis gives a value of $\gamma^*$ which is slightly different from the mean-field result. We cannot however benchmark our results with other methods, so it is difficult to  determine if the difference is significant. Notice that although both the dissipative TFI and XYZ models break the $\mathbb{Z}_2$ symmetry through the phase transition, the critical exponents are discriminated.

\begin{figure}[htb]
  \includegraphics[width=1\linewidth]{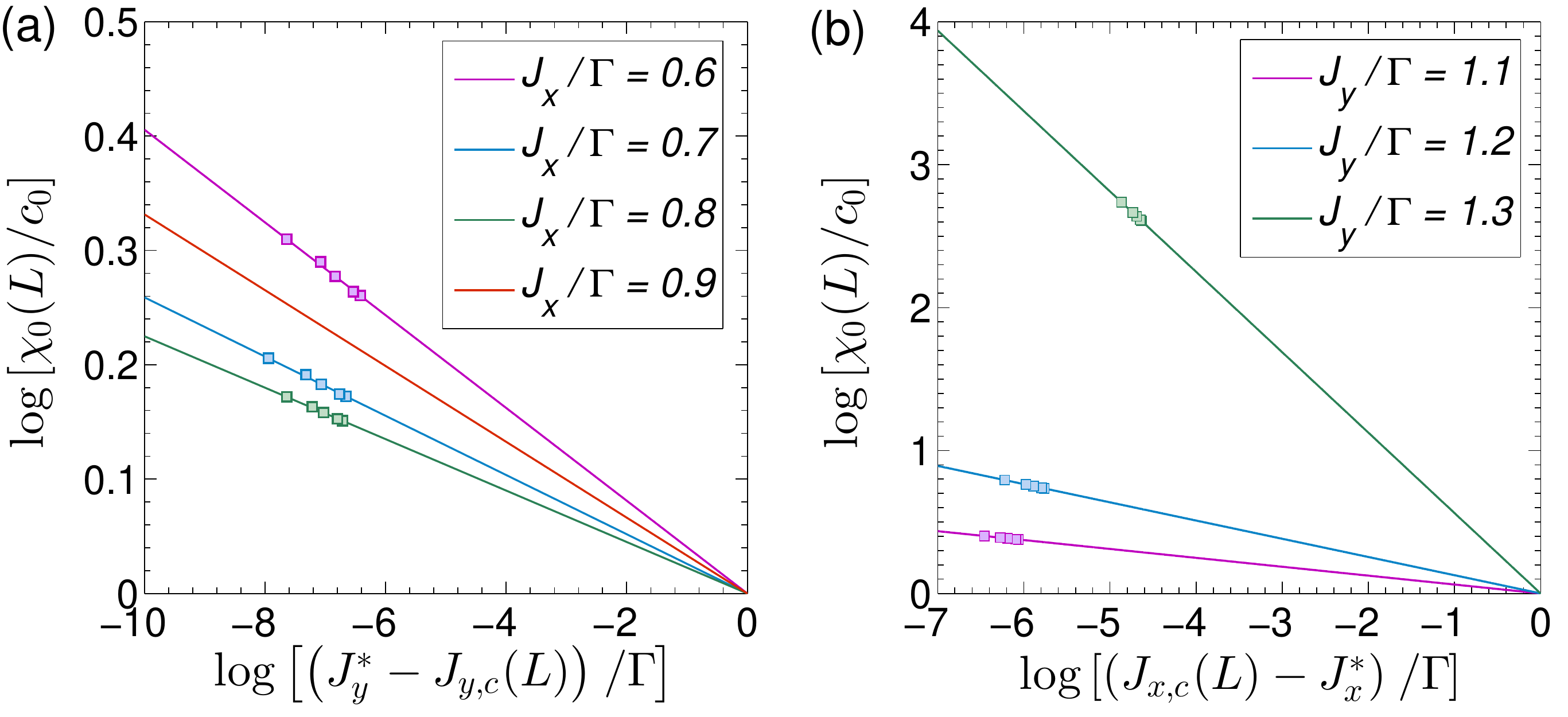}
  \caption{(a) The amplitudes $\chi_0(L)$ as functions of $\left[J_c(L)-J^*\right]/\Gamma$ for different values of (a) $J_x/\Gamma$ and (b) $J_y/\Gamma$. The squares denote the amplitudes for each of the clusters while the line denote the fittings with the critical exponents extracted from the CAM analysis. We have put the $\chi_0(L)$ for anisotropic clusters in the plots. From the left to right, the squares correspond to the $1\times1$, $1\times2$, $2\times2$, $2\times3$, and $3\times3$ clusters, respectively. The coupling is chosen as $J_z/\Gamma = 1$.
  }
  \label{camJxJy}
\end{figure}

\begin{table*}
\centering
\caption{The coefficients, critical points and critical exponents of steady-state phase transition in the dissipative TFI model extracted from the CAM with different series of clusters.}
\begin{tabular}{p{1.2cm}|p{1.8cm}p{1.8cm}p{1.8cm}|p{1.8cm}p{1.4cm}p{1.8cm}|p{1.8cm}p{1.4cm}p{1.8cm}cccccccccc}
\hline
\hline
&\multicolumn{3}{c|}{($1\times1$, $2\times2$, $3\times3$, $4\times4$)} & \multicolumn{3}{c|}{($1\times1$, $2\times2$, $3\times3$)} &\multicolumn{3}{c}{($2\times2$, $3\times3$, $4\times4$)} \\
\hline
$J_x/\Gamma$&$c_0$&$J_y^*/\Gamma$&$\gamma^*$&$c_0$&$J_y^*/\Gamma$&$\gamma^*$&$c_0$&$J_y^*/\Gamma$&$\gamma^*$ \\
\hline
0.9&0.121&1.041&1.033 &0.125&1.041&1.030&0.121&1.041&1.034\\
0.8&  - & - & - &0.0659& 1.021 & 1.023& -& -& -\\
0.7& - & - & - &   0.0433& 1.014 & 1.026& - & - & - \\
0.6 & - & - & - & 0.0298& 1.011 & 1.042 & - & - & - \\
\hline
\hline
\end{tabular}
\label{cam_XYZ1}
\end{table*}

Additionally, we also apply CAM analysis for $J_x/\Gamma=0.6$, $0.7$ and $0.8$ as well as $J_y/\Gamma=1.1$,  $1.2$,  and $1.3$ by using the $1\times1$, $2\times2$, and $3\times3$ clusters. The results are listed in Tabs. \ref{cam_XYZ1} and \ref{cam_XYZ2}. In Fig. \ref{camJxJy}, we show the amplitudes $\chi_0(L)$ as functions of $[J_{c}(L)-J^*]/\Gamma$ in the log-log scale. One can see that for case of fixed $J_x$, the critical exponents are stable around $\gamma^*\approx 1.03$, while for the case of fixed $J_y$, the critical exponents change in a wide range. We notice that for large $J_y$ the system is close to the critical regime in which the correlation length increases and clusters with larger size should be involved in the CAM analysis.

\begin{table}
\centering
\caption{The coefficients, critical points and critical exponents of steady-state phase transition in the dissipative XYZ model. The involved clusters are $1\times1$, $2\times2$, and $3\times3$.}
\begin{tabular}{p{2cm}p{2cm}p{2cm}p{2cm}}
\hline
\hline
 $J_y/\Gamma$ & $c_0$ & $J_x^*/\Gamma $ & $\gamma^*$ \\
 \hline
  1.1  & 0.112 & 0.959  & 1.062  \\
  1.2  & 0.0382&  0.978 & 1.128 \\
  1.3  & 0.00389& 0.977  & 1.563 \\
\hline
\hline
\end{tabular}
\label{cam_XYZ2}
\end{table}

\section{ Summary}
\label{sec_summary}
In this work we have introduced the CAM as an effective method to determine critical exponents of phase transitions in driven-dissipative systems. In particular, we have applied CAM to the dissipative TFI model and XYZ model in two dimensions, finding a critical exponent for the magnetic susceptibility which is different from the classical value. Our study indicates that CAM is powerful in accessing the critical properties, in comparison to the resources required. We notice that although the implementation of CAM in the parameter regions considered in this paper is valid, one should be careful when going outside these regions. For example, the continuous phase transition in dissipative TFI model may break down for sufficiently strong dissipation where the CAM analysis cannot be implemented \cite{overbeck2017}.

We would like to emphasize that the sizes of the involved clusters also have effects on the performance of CAM for open systems. In other words, for systems that are dominated by long spatial correlations, large size clusters should be involved in the CAM analysis, in order to reach reliable results.  For example, in the one-dimensional driven pair contact process with diffusion (PCPD) the critical exponents obtained through the CAM analysis are in excellent accord with simulation results \cite{park2005}. However, for ordinary PCPD (without driving) a non-negligible discrepancy is observed up to the maximum cluster size 13. This reminds us that the results obtained by CAM should be interpreted with caution.

To further enlarge the cluster size, there are several promising approaches worth to pursue in future work, involving a combination of corner-space renormalization~\cite{finazzi2015}, neural-networks~\cite{vicentini2019,hartmann2019,Yoshioka2019,Nagy2019} and/or tensor-networks~\cite{weimer2019,Paeckel2019}. On one side, the corner-space renormalization method and the recently introduced neural-networks quantum ansatz were shown to simulate the steady states of open quantum many-body systems, benchmarked with
a few examples including small $2$D clusters.  At the moment, it is not clear how large are the computational cost and convergence of the methods for:  (i) larger clusters sizes; (ii) quantum states close to criticality, for which one expects highly-entangled states; (iii) their extension to time-dependent open dynamics (as required in CAM). On the other side, tensor-networks approaches are well established methods with good convergence for short-range systems and low-entangled states. The extension of the method to $2$D clusters (effectively mapped to a long-range $1$D system) turns the simulation computationally harder at the expense of an increasing memory cost and simulation time. One could still exploit the benefits of CAM in this case, recalling that it requires a large but not \textit{macroscopically} large cluster, such that high performance supercomputers with multiple nodes may hopefully lead to an efficient simulation.

\section*{ACKNOWLEDGMENTS}
We thank M. F. Maghrebi and J. Keeling for useful correspondence. J.J. acknowledges support from the National Natural Science Foundation of China (NSFC) via Grant No. 11975064. W.B.He acknowledges support from Grant No. U1930402 of the National Science Association Funds of NSFC. F.I. acknowledges the financial support of the Brazilian funding agencies National Council for Scientific and Technological Development - CNPq (Grant No. $308205/2019-7$) and FAPERJ (Grant No. E-$26/211.318/2019$). Y.-D. W. acknowledges support by the National Key R\&D Program of China under Grant No. 2017YFA0304503, and the Peng Huanwu Theoretical Physics Renovation Center under grant No. 12047503. S.C. acknowledges support from the National Key R\&D Program of China No. 2016YFA0301200 and the NSFC Grant No. 11974040. R. F. acknowledges partial financial support from the Google Quantum Research Award. R. F. research has been conducted within the framework of the Trieste Institute for Theoretical Quantum Technologies (TQT).

\end{document}